\documentclass[a4paper]{article}

\usepackage{INTERSPEECH2021}

\usepackage{graphicx}
\usepackage{dcolumn}
\usepackage{amsmath,amsfonts}
\usepackage{bm}
\usepackage{multirow}
\usepackage{slashbox}

\title{Siamese Neural Network with Joint Bayesian Model Structure for Speaker Verification}
\name{Xugang Lu$^{1*}$, Peng Shen$^{1}$, Yu Tsao$^{2}$, Hisashi Kawai$^1$}
\address{
  $^1$National Institute of Information and Communications
Technology, Japan\\
  $^2$Research Center for Information Technology Innovation, Academic
Sinica, Taiwan
\thanks{The work is partially supported by JSPS KAKENHI
No. 19K12035}}
\email{xugang.lu@nict.go.jp}

\begin{document}

\maketitle
\begin{abstract}
Generative probability models are widely used for speaker verification (SV). However, the generative models are lack of discriminative feature selection ability. As a hypothesis test, the SV can be regarded as a binary classification task which can be designed as a Siamese neural network (SiamNN) with discriminative training. However, in most of the discriminative training for SiamNN, only the distribution of pair-wised sample distances is considered, and the additional discriminative information in joint distribution of samples is ignored. In this paper, we propose a novel SiamNN with consideration of the joint distribution of samples. The joint distribution of samples is first formulated based on a joint Bayesian (JB) based generative model, then a SiamNN is designed with dense layers to approximate the factorized affine transforms as used in the JB model. By initializing the SiamNN with the learned model parameters of the JB model, we further train the model parameters with the pair-wised samples as a binary discrimination task for SV. We carried out SV experiments on data corpus of speakers in the wild (SITW) and VoxCeleb. Experimental results showed that our proposed model improved the performance with a large margin compared with state of the art models for SV.
\end{abstract}

\section{Introduction}
\label{sec-I}
Speaker verification (SV) is a task to judge whether an utterance is spoken by a registered speaker or not, it is widely used in many speech application systems for authentic or security purpose \cite{Hansen2015,Poddar2018,Beigi2007}. The conventional pipeline in constructing a SV system is composed of a front-end speaker embedding feature extraction and a back-end speaker classifier modeling. The front-end embedding feature extraction tries to extract robust and discriminative speaker features, and back-end classifier tries to model speaker features based on which the similarity or distance between two compared features vectors could be estimated. In most state of the art frameworks, i-vector \cite{Dehak2011,Snyder2018}, d-vector and X-vector \cite{Snyder2018,Variani2014}, have been proposed as front-end speaker features. Particularly, the X-vector as one of the speaker embedding representations is the most widely used one \cite{Snyder2018}. Since the original front-end feature encodes various of acoustic factors, e.g., speaker factor, channel transmission factor, recording device factor, etc., before classifier modeling, a linear discriminative analysis (LDA) based dimension reduction is usually applied to eliminate non-speaker specific information. Based on the robust speaker features, several back-end speaker models have been proposed, for example, the probabilistic linear discriminant analysis (PLDA) modeling \cite{Dehak2011,Prince2007}, joint Bayesian (JB) modelling \cite{ChenJB2012,ChenJB2016}, support vector machine (SVM) \cite{WANSVM2000}, as well as other types of discriminative classification based modeling  \cite{Villalba2017} \cite{Burget2011,Cumanni2013}.

The SV problem can be defined as a hypothesis test \cite{Lehmann2005}:
\begin{equation}
\begin{array}{l}
H_S :{\bf x}_i ,{\bf x}_j \mathop {}\limits^{} \text{are spoken by the same speaker} \\
H_D :{\bf x}_i ,{\bf x}_j \mathop {}\limits^{} \text{are spoken by different speakers},\\
\end{array}
\label{eq_HSD}
 \end{equation}
where $H_S$ and $H_D$ are the two hypothesises as the same and different speaker spaces, respectively. $({\bf x}_i ,{\bf x}_j$) is a tuple with two compared utterances indexed by $i$ and $j$ (as a trial in SV tasks). In most of the SV algorithms, the hypothesis test defined in Eq. (\ref{eq_HSD}) is finally formulated as a log-likelihood ratio (L\_LLR) function \cite{Lehmann2005}. And usually the L\_LLR is estimated based on generative probabilistic models in a transformed speaker feature space. However, the generative models are lack of discriminative feature selection ability, and usually a discriminative feature transform is independently applied before the generative classifier modeling. As an alternative, the hypothesis test in Eq. (\ref{eq_HSD}) can be regarded as a binary classification task where neural network based discriminative models could be applied. In most of these discriminative models, a distance metric could be learned for the hypothesis test defined in Eq. (\ref{eq_HSD}). In this distance metric learning, no probability distribution assumption (e.g., Gaussian for most generative models) is required, and the feature transformed space and hypothesis test model can be optimized in a unified neural network model. However, in most of these distance metric learning algorithms, only the distribution of the distances of pair-wised samples is considered without considering the joint distribution of samples. As indicated in a joint Bayesian (JB) analysis model, considering the joint distribution of samples could introduce additional discriminative information compared with only considering the distribution of distances of the pair-wised samples \cite{ChenJB2012,ChenJB2016}. In this paper, we propose a novel Siamese neural network (SiamNN) based discriminative framework for SV. In the framework, the SiamNN architecture is designed to integrate the model structure of JB, and jointly optimized with a discriminative feature learning process. Due to the discriminative learning property, a direct evaluation metric for SV is easily integrated as a learning objective function. Our experiments confirmed the advantages of the proposed SiamNN framework.
\section{The proposed discriminative neural network model}
\label{sec-II}
The hypothesis test defined in Eq. (\ref{eq_HSD}) can be regarded as a Bayesian binary classification task, and a discriminative neural network model can be designed for the task. During the architecture design of the neural network model, in order to take the model structure of a generative probabilistic model into consideration, we need to explain the conventional generative model based algorithms, and their connections to the discriminative framework via the L\_LLR estimation.
\subsection{Log-likelihood ratio function based on generative probabilistic models}
Based on a generative probability model, a log-likelihood ratio (L\_LLR) with consideration of intra-speaker and inter-speaker distances is defined as:
\begin{equation}
r_{i,j} = r({\bf x}_i ,{\bf x}_j ) = \log \frac{{p(\Delta _{i,j} |H_S )}}{{p(\Delta _{i,j} |H_D )}},
\label{eq_bayesian}
\end{equation}
where $\Delta _{i,j}  = {\bf x}_i  - {\bf x}_j$ is the pair-wised sample distance. Based on the Gaussian density distribution assumptions of $p(. |H_S )$ and $p(. |H_D )$ , the L\_LLR can be estimated as:
\begin{equation}
r_{i,j}  = ({\bf x}_i  - {\bf x}_j )^T {\bf M}({\bf x}_i  - {\bf x}_j )
\label{eq_MD}
\end{equation}
where ${\bf M} = -(\Sigma _{H_S }^{ - 1}  - \Sigma _{H_D }^{ - 1}$) with $\Sigma _{H_S }$ and $\Sigma _{H_D }$ as the covariance matrices of the pair-wised distance space $\Delta _{i,j}$ for $H_S$ and $H_D$ conditions, respectively. We can see that Eq. (\ref{eq_MD}) is in the same form as  Mahalanobis distance metric except the negativity of the $\bf M$ \cite{Xing2002}.

From the definition in Eq. (\ref{eq_bayesian}), we can see that the learned distance metric only considers the distribution of the pair-wised sample distance space \cite{BayesianFace}. For joint Bayesian (JB) probability distribution modeling, i.e., ${p\left( {{\bf x}_i ,{\bf x}_j |H_S } \right)}$ or ${p\left( {{\bf x}_i ,{\bf x}_j |H_D } \right)}$, the L\_LLR is estimated as:
 \begin{equation}
 r_{i,j} = r({\bf x}_i ,{\bf x}_j ) = \log \frac{{p({\bf x}_{i,} {\bf x}_j |H_S )}}{{p({\bf x}_{i,} {\bf x}_j |H_D )}}
 \label{eq_lllr}
 \end{equation}
In the JB modeling, the observed speaker feature variable ${\bf x}$ satisfies the following formulation as:
\begin{equation}
{\bf x} = {\bf u}  + {\bf n},
\label{eq_add}
\end{equation}
where $\bf u$ is a speaker identity vector variable, and $\bf n$ represents intra-speaker variation caused by noise. In verification, for given a trial with ${\bf x}_i $ and ${\bf x}_j $ generated from Eq. (\ref{eq_add}), with zero mean Gaussian assumption (with covariance matrix $\Sigma _{\bf u}$ and $\Sigma _{\bf n}$  for $\bf u$ and $\bf n$ variables, respectively), the two terms ${p\left( {{\bf x}_i ,{\bf x}_j |H_S } \right)}$ and ${p\left( {{\bf x}_i ,{\bf x}_j |H_D } \right)}$ defined in Eq. (\ref{eq_lllr}) satisfy zero-mean Gaussian with covariances as:
\begin{equation}
\begin{array}{l}
 {\bf S}_{H_S }  = \left[ {\begin{array}{*{20}c}
   {\Sigma _{\bf u}  + \Sigma _{\bf n} } & {\Sigma _{\bf u} }  \\
   {\Sigma _{\bf u} } & {\Sigma _{\bf u}  + \Sigma _{\bf n} }  \\
\end{array}} \right] \\
 {\bf S}_{H_D }  = \left[ {\begin{array}{*{20}c}
   {\Sigma _{\bf u}  + \Sigma _{\bf n} } & {\bf 0}  \\
   {\bf 0} & {\Sigma _{\bf u}  + \Sigma _{\bf n} }  \\
\end{array}} \right] \\
 \end{array}
\label{eq_covJB}
\end{equation}
Based on Eq. (\ref{eq_covJB}), the L\_LLR defined in Eq. (\ref{eq_lllr}) could be calculated based on:
 \begin{equation}
 r( {{\bf x}_i ,{\bf x}_j } ) = {\bf x}_i^T {\bf Ax}_i  + {\bf x}_j^T {\bf Ax}_j  - 2{\bf x}_i^T {\bf Gx}_j,
 \label{eq_rij}
 \end{equation}
where
{\small
\begin{equation}
\begin{array}{l}
 {\bf A} = (\Sigma _{\bf u}  + \Sigma _{\bf n} )^{ - 1}  - [(\Sigma _{\bf u}  + \Sigma _{\bf n} ) - \Sigma _{\bf u} (\Sigma _{\bf u}  + \Sigma _{\bf n} )^{ - 1} \Sigma _{\bf u} ]^{ - 1}  \\
 {\bf G} =  - (2\Sigma _{\bf u}  + \Sigma _{\bf n} )^{ - 1} \Sigma _{\bf u} \Sigma _{\bf n}^{ - 1}  \\
 \end{array}
 \label{eq_jbAG}
\end{equation}
}
Comparing Eqs. (\ref{eq_MD}) and (\ref{eq_rij}), we can see that if we set ${\bf A}={\bf G}={\bf M}$, the JB model based L\_LLR degenerates to be the same form as the Mahalanobis distance metric (except the negativity of the matrix). In this sense, we can regard the L\_LLR in Eq. (\ref{eq_MD}) as a special case in JB model based estimation. Since the L\_LLR in Eqs. (\ref{eq_MD}) and (\ref{eq_rij}) are based on probabilistic modeling with Gaussian distribution assumptions, their model parameters could be estimated using EM (or EM-like) learning algorithms \cite{ChenJB2012, ChenJB2016}.
\subsection{Connecting log-likelihood ratio in a neural network classification model}
The L\_LLR defined either in Eqs. (\ref{eq_bayesian}) or (\ref{eq_lllr}) can be derived from a generative model based classification model. Given a training data set $\left\{ {\left( {{\bf x}_i ,y_i } \right)} \right\}_{i = 1,2,...,N} ,y_i  \in \left\{ {1,2,...,K} \right\}$ with ${\bf x}_i$ and $y_i $ as data feature and label, $K$ is the number of classes, the classification model is defined as:
\begin{equation}
p\left( {y = k|{\bf x}} \right) = \frac{{p\left( {{\bf x}|y = k} \right)p\left( {y = k} \right)}}{{\sum\limits_{j = 1}^K {p\left( {{\bf x}|y = j} \right)p\left( {y = j} \right)} }}.
\label{eq_Gen1}
\end{equation}
And Eq. (\ref{eq_Gen1}) is further cast to:
\begin{equation}
p\left( {y = k|{\bf x}} \right) = \frac{1}{{1 + \sum\limits_{j = 1,j \ne k}^K {\exp \left( {-r_{k,j} \left( {{\bf x},\Theta } \right)} \right)} }},
\label{eq_Gen2}
\end{equation}
where
\begin{equation}
r_{k,j} \left( {{\bf x},\Theta } \right) \mathop  = \limits^\Delta \log \frac{{p\left( {{\bf x}|y = k} \right)p\left( {y =k} \right)}}{{p\left( {{\bf x}|y = j} \right)p\left( {y = j} \right)}},
\label{eq_GenG}
\end{equation}
is a L\_LLR function based on the probabilistic model with $\Theta $ as a model parameter set. In a neural network based classification model, the classification is formulated as:
\begin{equation}
p\left( {y = k|{\bf x}} \right) = \frac{{\exp \left( {o_k } \right)}}{{\sum\limits_{j = 1}^K {\exp \left( {o_j } \right)} }},
\label{eq_softmax}
\vspace{-2mm}
\end{equation}
where a network mapping function $o_j  = \phi _j \left( {{\bf x},\Theta } \right)$ is defined as the output corresponding to the $j$-th class, and $\Theta$ is the neural network parameter set. And Eq. (\ref{eq_softmax}) is cast to:
\begin{equation}
p\left( {y = k|{\bf x}} \right) = \frac{1}{{1 + \sum\limits_{j = 1,j \ne k}^K {\exp \left( {-h_{k,j} \left( {{\bf x},\Theta } \right)} \right)} }},
\label{eq_Dist}
\vspace{-2mm}
\end{equation}
where
\begin{equation}
h_{k,j} \left( {{\bf x},\Theta } \right) \mathop  = \limits^\Delta \phi _k \left( {{\bf x},\Theta} \right) - \phi _j \left( {{\bf x},\Theta} \right).
\label{eq_DistD}
\end{equation}
Comparing Eqs. (\ref{eq_Dist}), (\ref{eq_DistD}) with (\ref{eq_Gen2}), (\ref{eq_GenG}), we can see that $h_{k,j} \left( {{\bf x},\Theta} \right)$ can be connected to the L\_LLR in calculation in a pair-wised neural discriminative training.
\subsection{Pair-wised discriminative training for L\_LLR modeling}
For convenience of formulation, we define a trial as a tuple ${\bf{ z}}_{i,j}  = ( {{\bf x}_i ,{\bf x}_j } )$, and the two hypothesis spaces are constructed from the two data sets as:
\begin{equation}
\begin{array}{l}
 S = \left\{ {{\bf{z}}_{i,j}  = ( {{\bf x}_i ,{\bf x}_j } ) \in H_S } \right\} \\
 D = \left\{ {{\bf{z}}_{i,j}  = ( {{\bf x}_i ,{\bf x}_j } ) \in H_D } \right\} \\
 \end{array}
\label{eq_set}
\end{equation}
Given a trial with two observation variables ${\bf z}_{i,j}  = ({\bf x}_i {\rm ,}{\bf x}_j )$ (X-vectors in this study), the classification task is to estimate and compare $p(H_S |{\bf z}_{i,j} )$ and $p(H_D |{\bf z}_{i,j} )$. As a binary discriminative learning, the label is defined as:
\begin{equation}
y_{i,j}  = \left\{ {\begin{array}{*{20}c}
   { 1,{\bf z}_{i,j}  \in H_S }  \\
   { 0,{\bf z}_{i,j}  \in H_D }  \\
\end{array}} \right.
 \label{eq_H1H0}
\end{equation}
Based on discriminative neural network model with reference to Eqs. (\ref{eq_Dist}) and (\ref{eq_DistD}), the posterior probability is estimated based on:
\begin{equation}
p(y_{i,j} |{\bf z}_{i,j} ) = \left\{ \begin{array}{l}
 \frac{1}{{1 + \exp ( - h_{H_S ,H_D } ({\bf z}_{i,j} ,\Theta ))}};{\bf z}_{i,j}  \in H_S  \\
 1 - \frac{1}{{1 + \exp ( - h_{H_S ,H_D } ({\bf z}_{i,j} ,\Theta ))}};{\bf z}_{i,j}  \in H_D  \\
 \end{array} \right.
\label{eq_DisB}
\end{equation}
As we have revealed from Eqs. (\ref{eq_Gen2}), (\ref{eq_GenG}), and (\ref{eq_lllr}), we replace the ${h_{H_S ,H_D } ({\bf z}_{i,j} ,\Theta )}$ with L\_LLR function, and define a mapping as a logistic function with scaled parameters as \cite{Platt1999,HLin2007}:
\begin{equation}
f\left( {r_{i,j} } \right)\mathop  = \limits^\Delta  \frac{1}{{1 + \exp \left( { - \left( {\alpha  r_{i,j}  + \beta } \right)} \right)}}
\label{eq_JointDG}
\end{equation}
where $r_{i,j} $ is the L\_LLR as defined in either Eq. (\ref{eq_bayesian}) or (\ref{eq_lllr}), $\alpha$ and $\beta$ are gain and bias factors used in the regression model. In Eq. (\ref{eq_JointDG}), we integrate the L\_LLR score estimated from the probabilistic model in a neural discriminative training framework. The probability estimation in Eq. (\ref{eq_DisB}) is cast to:
\begin{equation}
\hat y_{i,j} \mathop  = \limits^\Delta  p(y_{i,j} |{\bf z}_{i,j} ) = \left\{ {\begin{array}{*{20}c}
   {f(r_{i,j} );{\bf z}_{i,j}  \in H_S }  \\
   {1 - f(r_{i,j} );{\bf z}_{i,j}  \in H_D }  \\
\end{array}} \right.
\end{equation}
The model parameters can be learned based on optimizing binary classification accuracy. Under this framework, it is easy to directly incorporate the SV evaluation metric in the neural discriminative leaning. In this study, an empirical Bayes risk (EBR) based objective function is adopted with consideration of the false alarm and miss detections, which is widely used in hypothesis test tasks for SV\cite{BOSARIS2011,SITW2016}.
\subsection{Integrating the generative probabilistic model structure in the discriminative neural network}
\begin{figure}[tb]
\centering
\includegraphics[width=7cm, height=5.8cm]{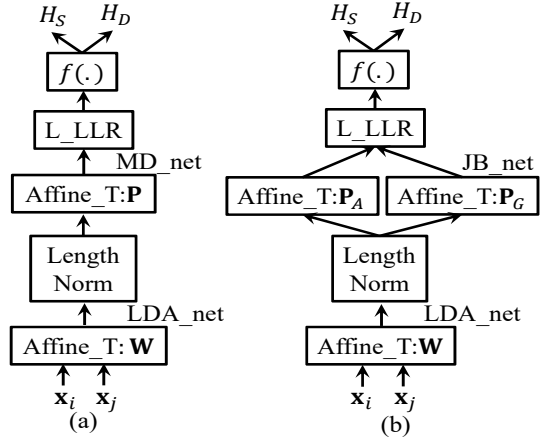}
\caption{The proposed SiamNN framework with: (a) MD\_net (Mahalanobis net) with an affine transform matrix $P$, and (b) JB\_net (joint Bayesian net) with two branches of affine transform matrices $P_A$ and $P_G$. LDA\_net with an affine transform matrix $W$ for fulfilling the LDA transform.}
\label{figNJBNet}
\vspace{-2mm}
\end{figure}

We design a Siamese neural network (SiamNN) within a pair-wised discriminative learning framework for SV.
In conventional pipeline for SV, a LDA is applied before the probabilistic modeling. Correspondingly, in the SiamNN, a dense layer is designed for fulfilling the function of LDA, and another dense layer is for fulfilling the transform functions used in Eqs. (\ref{eq_MD}) and (\ref{eq_rij}). For more specific, the transform matrix used in Eq. (\ref{eq_MD}) is factorized as:
\begin{equation}
{\bf M} = -{\bf PP}^T.
\end{equation}
And the matrices in Eq. (\ref{eq_rij}) are factorized as:
\begin{equation}
\begin{array}{l}
 {\bf A} =  - {\bf P}_A {\bf P}_A^T  \\
 {\bf G} =  - {\bf P}_G {\bf P}_G^T  \\
 \label{eq_AGF}
 \end{array}
\end{equation}
Based on the factorizations, the L\_LLR function in Eq. (\ref{eq_MD}) is cast to:
\begin{equation}
r_{i,j}  = 2b_i^T b_j  - b_i^T b_i  - b_j^T b_j
\end{equation}
with affine transforms as:
\begin{equation}
 b_k  = {\bf P}^T {\bf \tilde h}_k.
 \label{eq_P}
 \end{equation}
And the L\_LLR function in Eq. (\ref{eq_rij}) is cast to:
\begin{equation}
r_{i,j} = 2g_i^T g_j  - a_i^T a_i  - a_j^T a_j
\label{eq_Dlllr}
\end{equation}
with the affine transforms as:
\begin{equation}
\begin{array}{l}
 a_i  = {\bf P}_A^T {\bf \tilde h}_i  \\
 g_i  = {\bf P}_G^T {\bf \tilde h}_i,\\
 \label{eq_PAGF}
 \end{array}
 \vspace{-2mm}
\end{equation}
where in Eqs. (\ref{eq_P}) and (\ref{eq_PAGF}), $k \in \{ i,j\}$, ${\bf \tilde h}_i  = \frac{{{\bf h}_i }}{{||{\bf h}_i ||}}$ is the length normalized vector from the LDA transform as:
 \begin{equation}
 {\bf h}_i  = {\bf W}^T {\bf x}_i,
 \label{eq_LDA}
 \end{equation}
where ${\bf x}_i$ is the input X-vector feature, $\bf W$ is the transform in LDA. With these factorizations, the model architecture is designed as illustrated in Fig. \ref{figNJBNet}.
In this figure, "LDA\_net" is for the LDA transform with the affine transform "Affine\_T" defined in Eq. (\ref{eq_LDA}), "MD\_net" is the Mahalanobis distance net with affine transform defined in Eq. (\ref{eq_P}), and "JB\_net" is the JB network with two branches of affine transforms defined in Eq. (\ref{eq_PAGF}).
\section{Experiments and results}
\label{sec-III}
\subsection{Experimental conditions}
We carried out SV experiments to test our proposed framework. The training data set is from VoxCeleb data corpus (sets 1 and 2) \cite{Vox2020}, and the test data sets are from the data corpus of speakers in the wild (SITW) \cite{SITW2016}. We adopt a state of the art pipeline for constructing the SV baseline systems. The input speaker feature in our pipeline is X-vector which is extracted from a well trained deep time delay neural network (TDNN) neural network \cite{Snyder2018}. In training the TDNN model, the training data includes two data sets from Voxceleb corpus, i.e., the training set of Voxceleb1 corpus by removing overlapped speakers which are included in the test set of the SITW, and the training set of Voxceleb2. Moreover, data augmentation is applied to increase data diversity in TDNN model training. Input features for training the speaker embedding model are 30 Mel band bins based MFCCs with 25 ms frame length and 10 ms frame shift. The final extracted X-vector is with 512 dimensions. For removing non-speaker information, the LDA is applied to transform the 512-dimension X-vectors to 200-dimension vectors before the probabilistic modeling. Correspondingly, in the discriminative neural network model as showed in Fig. \ref{figNJBNet}, a dense layer with 200 neurons is also applied in the "LDA\_net".

Since the discriminative neural network architecture fits well to the conventional pipeline based on the probabilistic model structure, the dense layer parameters could be initialized with the conventional model parameters in training (according to Eqs. (\ref{eq_LDA}) and (\ref{eq_PAGF})). For comparison, the random parameter initialization method is also examined. In model training, the Adam algorithm with an initial learning rate of $0.0005$ \cite{Adam} was used. In order to include enough "negative" and "positive" samples, the mini-batch size was set to 4096. The training X-vectors were splitted to training and validation sets with a ratio of $9:1$. The model parameters were selected based on the best performance on the validation set.
\subsection{Results}
Two testing data sets from the SITW, i.e., development and evaluation sets are used, and each is used as an independent test set. The evaluation metrics, equal error rate (EER) and minimum decision cost function (minDCF) (with target prior 0.01 denoted as minDCF1, and prior 0.001 denoted as minDCF2) are adopted to measure the performance \cite{BOSARIS2011,SITW2016}. The results are showed in tables (\ref{tab1}) and (\ref{tab2}).
\begin{table}[tb]
\centering
\caption{Performance on the development set of SITW.}
\vspace{-2mm}
\begin{tabular}{|c||c|c|c|}
\hline
 Methods&EER(\%)&minDCF1&minDCF2\\
\hline
\hline
LDA+PLDA  &3.00&0.332&0.520\\
\hline
LDA+JB  &3.04&0.329&0.502\\
\hline
SiamNN (rand init)  &4.16&0.379&0.588\\
\hline
SiamNN (JB init) &\textbf{2.66}&\textbf{0.297}&\textbf{0.447}\\
\hline
\end{tabular}
\label{tab1}
\end{table}
\begin{table}[tb]
\centering
\vspace{-2mm}
\caption{Performance on the evaluation set of SITW.}
\begin{tabular}{|c||c|c|c|}
\hline
 Methods&EER (\%)&minDCF1&minDCF2\\
\hline
\hline
LDA+PLDA  &3.55&0.353&0.566\\
\hline
LDA+JB  &3.50&0.342&0.565\\
\hline
SiamNN (rand init)  &4.51&0.392&0.600\\
\hline
SiamNN (JB init)  &\textbf{3.14}&\textbf{0.308}&\textbf{0.462}\\
\hline
\end{tabular}
\label{tab2}
\vspace{-2mm}
\end{table}
In these two tables, "LDA+PLDA" and "LDA+JB" represent the baseline systems based on probabilistic models PLDA and JB, respectively.   "SiamNN" denotes the proposed system which takes the probabilistic JB model structure in designing the neural network model (as illustrated in (b) of Fig. \ref{figNJBNet}), and model parameters are with random initialization ("SiamNN (rand init)") or with EM algorithm learned JB model parameters ("SiamNN (JB init)"). From these two tables, we can see that the performance of the baseline system with probabilistic JB model is comparable or a slight better than that of the PLDA based model. In the SiamNN based model, if model parameters are randomly initialized ("SiamNN (rand init)"), the performance is worse than the original baseline model based results. However, when the SiamNN parameters are initialized with the JB based baseline model parameters, the performance is significantly improved. These results indicate that the discriminative training could further enhance the discriminative power of the conventional JB based probabilistic model.
\begin{table}[tb]
\centering
\caption{Performance before the SiamNN discriminative training (with JB init) on the development set of SITW.}
\begin{tabular}{|c||c|c|c|}
\hline
 Methods&EER (\%)&minDCF1&minDCF2\\
\hline
\hline
A (G=0)  &47.71&1.00&1.00\\
\hline
G (A=0) &6.35&0.826&0.981\\
\hline
A, G (set G to A) &\textbf{3.12}&\textbf{0.360}&\textbf{0.584}\\
\hline
A, G (set A to G) &3.50&0.398&0.632\\
\hline
\end{tabular}
\label{tab3}
\vspace{-2mm}
\end{table}
\begin{table}[tb]
\centering
\caption{Performance after the SiamNN discriminative training (with JB init) on the development set of SITW.}
\vspace{-2mm}
\begin{tabular}{|c||c|c|c|}
\hline
 Methods&EER (\%)&minDCF1&minDCF2\\
\hline
\hline
A (G=0)  &50.29&1.000&1.000\\
\hline
G (A=0)  &4.78&0.421&0.634\\
\hline
A, G (set G to A)  &\textbf{2.81}&\textbf{0.298}&0.456\\
\hline
A, G (set A to G)  &3.08&0.313&\textbf{0.451}\\
\hline
\end{tabular}
\label{tab4}
\end{table}
\begin{table}[tb]
\centering
\caption{Performance of the SiamNN discriminative training with "MD\_net" on the development set of SITW}
\vspace{-2mm}
\begin{tabular}{|c||c|c|c|}
\hline
 Methods&EER (\%)&minDCF1&minDCF2\\
\hline
\hline
Random init $\bf P$ &3.97&0.374&0.554\\
\hline
Init ${\bf P}$ with ${\bf P}_A$  &\textbf{3.62}&\textbf{0.369}&\textbf{0.547}\\
\hline
Init ${\bf P}$ with ${\bf P}_G$  &4.01&0.406&0.600\\
\hline
\end{tabular}
\label{tab_dm}
\vspace{-2mm}
\end{table}
\subsection{Effect of $\bf A$ and $\bf G$ on SV performance}
\label{sect_AG_effect}
In our SiamNN discriminative training, the L\_LLR of the JB model defined in Eq. (\ref{eq_rij}) is integrated. With different settings of $\bf A$ and $\bf G$ in Eq. (\ref{eq_rij}), we could obtain:
\begin{equation}
r({\bf x}_i ,{\bf x}_j ) = \left\{ {\begin{array}{*{20}c}
   { - 2{\bf x}_i^T {\bf Gx}_j ;\text{ for }{\bf A} = 0}  \\
   {{\bf x}_i^T {\bf Ax}_i  + {\bf x}_j^T {\bf Ax}_j ;\text{ for }{\bf G} = 0}  \\
   {({\bf x}_i  - {\bf x}_j )^T {\bf G}({\bf x}_i  - {\bf x}_j );\text{ for }{\bf A} = {\bf G}}  \\
   {({\bf x}_i  - {\bf x}_j )^T {\bf A}({\bf x}_i  - {\bf x}_j );\text{ for }{\bf G} = {\bf A}}  \\
\end{array}} \right.
\label{eq_maha_convert}
\vspace{-2mm}
\end{equation}
Based on this formulation, the two matrices $\bf A$ and $\bf G$  are connected to the two dense layers of the SiamNN model with weights ${\bf P}_A$ and ${\bf P}_G$ (refer to Fig. \ref{figNJBNet}). In our model, the dense layers were first initialized with the parameters from the learned JB based baseline model, then the model was further trained with "negative" and "positive" pair-wised samples. Only in testing stage, the different parameter settings according to Eq. (\ref{eq_maha_convert}) are examined for experiments, and the results are showed in tables \ref{tab3} and \ref{tab4} for the dev set of SITW. In these two tables, by comparing conditions with ${\bf A}=0$ or ${\bf G}=0$, we can see that the cross term contributes more to the SV performance, i.e., the dense layer with neural weight ${\bf P}_G$ contributes to the most discriminative information in the SV task. Moreover, when keeping the cross term either by setting ${\bf A} = {\bf G}$ or ${\bf G} = {\bf A}$, the performance is better than setting any one of them to be zero. As a special case of the JB model based SiamNN, we also test the "MD\_net" on dev set of SITW with different settings, and show the results in table \ref{tab_dm}. From this table, we can see that when the model parameters are initialized with the ${\bf P}_A$ parameters, the performance is the best for this "MD\_net" based model. However, no matter in what conditions, comparing results in tables \ref{tab1} and \ref{tab_dm}, we can confirm that the model structure inspired by the JB model is the best when the model parameters are initialized properly.
\section{Discussion and conclusion}
\label{sec-IV}
In this study, we regard SV problem as a Bayesian binary classification task, and propose a SiamNN discriminative learning framework with "positive" and "negative" sample pairs (as from the same and different speakers). Rather than only considering the distributions of pair-wised intra- and inter-speaker distances, the joint distribution of samples is taken into consideration via the formulation from JB based generative modelling. With the help of matrix factorization, we reformulate the L\_LLR estimation of the JB model to a distance metric as used in the discriminative learning framework. In particular, the linear transform matrices in the JB model are implemented as dense layers of the neural network model hence the JB based model structure is effectively connected to the SiamNN framework. Moreover, the SiamNN framework takes the speaker feature transform and classification model parameters learning in a unified optimization framework. Our experiments confirmed that the SV was benefitted from the unified discriminative learning framework. In this study, the JB model is based on a simple Gaussian distribution assumption of speaker features and noise. In real applications, the probability distributions are much more complex. Although it is difficult for a generative probabilistic model to fit complex probability distributions in a high dimensional space, it is relatively easy for a neural network learning framework to do it. In the future, we will consider model structures for dealing with more complex probability distributions in SV tasks.

\end{document}